\documentclass{jpp}

\usepackage[utf8]{inputenc}
\usepackage{latexsym}
\usepackage{graphics}
\usepackage{graphicx}
\usepackage{subcaption}
\usepackage{amsmath}
\usepackage{amssymb}
\usepackage{color}
\usepackage[T1]{fontenc}
\usepackage[english]{babel}
\usepackage{bm}
\shorttitle{On the polarization of shear Alfvén and acoustic continuous spectra}
\shortauthor{M.V. Falessi, N. Carlevaro, V. Fusco, E. Giovannozzi, G. Vlad, F. Zonca}

\title{On the polarization of shear Alfvén and acoustic continuous spectra in toroidal plasmas}

\author{M. V. Falessi\aff{1,2}
\corresp{\email{matteo.falessi@enea.it}},
N. Carlevaro\aff{1,3}, V. Fusco\aff{1}, E. Giovannozzi\aff{1},\\ P. Lauber\aff{4}, G. Vlad\aff{1}, F. Zonca\aff{1,5}}

\affiliation{
\aff{1}ENEA, Fusion and Nuclear Safety Department, C. R. Frascati,\\ Via E. Fermi 45, 00044 Frascati (Roma), Italy
\aff{2}INFN - Rome section, Piazz.le Aldo Moro 2, 00185 Roma, Italy
\aff{3}Consorzio RFX, Corso Stati Uniti 4, 35127 Padova, Italy
\aff{4} Max Planck Institute for Plasma Physics, 85748 Garching, Germany
\aff{5}Institute for Fusion Theory and Simulation and Department of Physics,\\ Zhejiang University, Hangzhou 310027, China
}

\begin{document}
\maketitle

\begin{abstract}
In this work, the \texttt{FALCON} code is adopted for illustrating the features of shear Alfvén and continuous spectra in toroidal fusion plasmas. The \texttt{FALCON} codes employs the local Floquet analysis discussed in [Phys. of Plasmas \textbf{26} (8), 082502 (2019)] for computing global structures of continuous spectra in general toroidal geometry. As particular applications, reference equilibria for the Divertor Tokamak Test and ASDEX Upgrade plasmas are considered. In particular, we illustrate the importance of mode polarization for recognizing the physical relevance of the various branches of the continuous spectra in the ideal MHD limit. We also analyze the effect of plasma compression and the validity of the \emph{slow sound approximation}.
\end{abstract}

\section{Introduction}
\label{sec:org014c842}
Fast ion destabilized Alfvén modes are of interest to Tokamaks, as they can lead to increased rapid ion losses and ultimately affect fusion efficiency. Therefore, the calculation of shear Alfvén wave (SAW) continuous spectrum \citep{barston64,sedlacek71,grad69,uberoi72,tataronis73,hasegawa74,chen74a,chen74b,dewar74,appert74,goedbloed75,pao75,chance77} is of crucial importance as it determines mode structures and dispersive properties of Alfvénic fluctuations excited by energetic particles (EPs) in fusion devices \citep{zonca14a,zonca14b}. It is well known that SAW and ion sound wave (ISW) continuous frequency spectra are coupled due to equilibrium magnetic field curvature \citep{pogutse78,dippolito80,kieras82,cheng85,cheng86} and that their structures should be considered self-consistently. Being able to describe realistic geometries and plasma non-uniformity is therefore of crucial importance to compare numerical results with experiments. For these reasons, following \citet{chen16,zonca14a,zonca14b,chen17}, in \citet{falessi2019shear} we have shown how to calculate the two frequency spectra by solving the MHD equations in the ballooning space describing SAW-ISW waves propagating along magnetic field lines in general toroidal geometry. This approach corresponds to a mathematical formulation in the framework of the Floquet theory of ordinary differential equations with periodic coefficients \citep{Salat_1997}; and it is equivalent to the typical one consisting in a spectral decomposition of the fluctuations  in poloidal and toroidal angles and the null space (kernel) computation of the matrix of the highest order radial derivative \citep{chance77}. Anyway, as already pointed out by \citet{chu92}, the Floquet approach \citep{falessi2019shear} simplifies the calculation of the convoluted continuous spectrum structures at high mode numbers and/or near the plasma boundary. Within this framework, the general description of SAWs and ISWs using Gyrokinetic theory has been given by \citet{chen16,zonca14a,zonca14b}, and allows to include thermal plasma \citep{zonca96b,zonca98,zonca99,lauber2013,bowden15,lauber2018} as well EPs \citep{chen2016physics,zonca14a,zonca14b}. Moreover, the generality of the formulation allows, in principle, the Floquet approach to be extended to include three-dimensional equilibrium effects. As shown by \citet{falessi2019shear}, there is a straightforward connection of the present approach with the calculation of the generalized plasma inertia in the general fishbone-like dispersion relation (GFLDR), that is, the unified framework for describing Alfvénic fluctuations excited by EPs in Tokamaks \citep{chen16,zonca14a,zonca14b}. The GFLDR can be used for macroscopic fluctuation stability and nonlinear dynamics studies to extract the distinctive features of the different Alfvén eigenmode (AE)/ Energetic Particle Mode (EPM) branches, thus illuminating the crucial physics responsible for the behaviors observed either experimentally or by numerical simulations. This is of fundamental importance because, despite the advances made in gyrokinetic simulations, the precise interpretation of the observed fluctuations is often not sufficiently clear and debatable. This is particularly the case of the mode structure and damping rate at the Beta induced Alfvén Acoustic Eigenmode (BAAE) frequency range which will be briefly commented in the next sections, keeping in mind the need of kinetic theory as pointed out in Refs. \citep{chavdarovski09,zonca10,chavdarovski14,zhang16,Bierwage2017,lauber2013}. As an additional advantage, the Floquet approach allows  to determine precise boundary conditions for the calculation of (radially) local parallel mode structures \citep{zonca14a,zonca14b}.

Below, we follow and systematize the findings by \citet{falessi2019shear}, that is a numerical approach for the calculation of coupled SAW-ISW Alfvén continuous spectra within the aforementioned theoretical framework, resulting in the development of the Floquet ALfvén CONtinuum code, i.e. \texttt{FALCON}. Thanks to the generality of the theoretical formulation, the code structure is extremely schematic and entirely written in \texttt{Python} without compromising code performances. While the work by \citet{falessi2019shear} was focused mainly on the calculation of continuum frequency, we now outline the important role of fluctuation polarization, for which we also propose the Alfvénicity parameter as a proxy to gain further insights and guide physics intuition about the qualitative importance of collisionless damping.  
More precisely, Alfvénicity can be used to characterize radial singular structures by superimposing its magnitude as color bar in the continuous spectrum plots allowing, at a glance, to determine the effective role of resonant excitation and to study the shaping and finite pressure effects on the continuous spectrum structures at low frequency \citep{chu92,huysmans95,goedbloed98,vanderholst00}. Below, SAW-ISW continuous spectra are calculated within three different levels of approximation for two reference cases, i.e. the ASDEX Upgrade (AUG) \(\#31213\) discharge and the Divertor Tokamak Test (DTT) double null scenario. The range of physical parameters of the two configurations, e.g. $\beta$, is large enough to investigate their effects on the continuum structures. Thus, our results allow
verification and validation of a given simplified model.

The paper is organized as follows. In Sec. \ref{sec:org67a2c18} we recall the theoretical framework introduced in \citet{falessi2019shear}. As original application, in Sec. \ref{sec:orge40f96e}-\ref{sec:org9b666dd}  we solve for the SAW and ISW spectra in the two aforementioned reference scenarios. In Sec. \ref{sec:org5e2748a}, we present our conclusions.
\section{Theoretical framework \& solution technique}
\label{sec:org67a2c18}
In the following Section, we briefly review the equations describing SAW-ISW continuum in the ideal MHD limit, which are solved by \texttt{FALCON} code using the mode structure decomposition approach \citep{lu12} that asymptotically reduces to the ballooning formalism. The general theoretical framework is explored in \citep{chen17} and is based on \citet{zonca14a,zonca14b} while, more recently, we have shown how Floquet theory \citep{floquet1883equations,hill1886part,magnus2013hill,Denk1995} can be applied to numerically solve this problem. 

In this work, we study Tokamak geometry and, therefore, without loss of generality, we assume an axisymmetric equilibrium magnetic field \(\boldsymbol{B}_{0}\) expressed in flux coordinates \((r,\theta,\varphi)\):
\begin{equation}
\label{eq:27}
\boldsymbol { B } _ { 0 } = F ( \psi ) \boldsymbol { \nabla } \varphi + \boldsymbol{\nabla} \varphi \times \boldsymbol{ \nabla } \psi\;,
\end{equation}
where $\psi$ is the poloidal magnetic flux, $r = r(\psi)$ is a radial-like flux coordinate,  \(\varphi\) is the physical toroidal angle and the poloidal-like angular coordinate \(\theta\) can be chosen such that the Jacobian \(J = ( \boldsymbol { \nabla } \psi \times \boldsymbol { \nabla } \theta \cdot \boldsymbol { \nabla } \varphi ) ^ { - 1 }\) has a convenient expression. Furthermore, we can define the straight field line toroidal angle $\zeta$ such that the safety factor
\begin{equation}
q = \frac{\bm B_0 \cdot \bm \nabla \zeta}{\bm B_0 \cdot \bm \nabla \theta} = q(r)
\end{equation}
is a flux function. Following \cite{Connor_1978}, we introduce the extended poloidal angle coordinate \(\vartheta\) and the following decomposition for the $n$-th toroidal harmonic of a scalar field $\Phi_s$:
\begin{eqnarray}
\label{eq:deco}  
\Phi_s (r,\theta,\zeta) & = & 2\pi \sum_{\ell \in \mathbb{Z}} e^{in\zeta-inq(\theta-2\pi\ell)}  \hat \Phi_s (r,\theta-2\pi\ell)= 
\nonumber \\ & = &  \sum_{m \in \mathbb{Z}} e^{in\zeta-i m\theta} \int_{- \infty}^{\infty} d\vartheta e^{i(m-nq)\vartheta} \hat \Phi_s (r,\vartheta)  \; .  \label{eq:msd}
\end{eqnarray}
In this expression, $\vartheta$ represents not only the ``extended poloidal angle'' along magnetic field lines, but also, due to finite magnetic shear, the effective expression of an adimensionalized radial wave vector. Radial singular structures corresponding to the continuous spectra are obtained from the limiting forms of vorticity and pressure equations for $|\vartheta|\rightarrow \infty$ \citep{zonca14a,zonca14b,chen17}. In the mapping to $\vartheta$ space, periodic poloidal angle dependences of equilibrium quantities are replaced by the same dependences on $\vartheta$. While fluctuations must be periodic in $\theta$ space for physical reasons, they are generally not periodic in $\vartheta$. In \cite{falessi2019shear} we have introduced an appropriate angular coordinate to express the SAW frequency continuum more transparently; i.e.:
\begin{equation}
\label{eq:1}
\eta(\vartheta) \equiv 2 \pi \frac{\int _ { 0 } ^ { \vartheta } d \vartheta ^ { \prime } |\bm \nabla r|^{-2}}{\int _ { 0 } ^ { 2 \pi } d \vartheta ^ { \prime } |\bm\nabla r|^{-2}}\;.
\end{equation}
By construction, the Jacobian \(J_{\eta}\) of this new set of coordinates, dubbed continuum flux coordinates (CFC), is such that \(J_{\eta} B_{0}^{2}/|\bm\nabla r|^{2}\) is a flux function. Following \citet{chen17,falessi2019shear}, we introduce the equations describing  SAW-ISW singular structures propagating along magnetic field lines in CFC coordinates:
\begin{align}
\label{eq:366}
\begin{split}
\left( \partial _ { \eta } ^ { 2 } + \hat { J } _ { \eta } ^ { 2 } \hat{\rho}_{m0} \Omega ^ { 2 } \right) g _ { 1 } &=  2  \hat{B} _ { 0 }   \hat{\rho}_{m0}^ { 1 / 2 } \hat { J } _ { \eta } ^ { 2 } \hat{\kappa} _ { g } \Omega\, g _ { 2 }\;, \\
\left( \partial _ { \eta } ^ { 2 } - |\bm \nabla r | \partial _ { \eta } ^ { 2 } | \bm\nabla r | ^ { - 1 } + \frac { 2 } { \Gamma \beta } \hat{\rho}_{m0} \hat { J } _ { \eta } ^ { 2 } \Omega ^ { 2 } \right) g _ { 2 } &=  2  \hat{B} _ { 0 }  \hat{\rho}_{m0}^ { 1 / 2 } \hat { J } _ { \eta } ^ { 2 } \hat{\kappa} _ { g } \Omega\, g _ { 1 } \;\\
\end{split}
\end{align}
where we have introduced the following dimensionless quantities:
\begin{equation}
\label{eq:33}
\Omega=\frac{\omega R_{0}}{\bar{v}_{A 0}}, \quad \hat{J}_{\eta}^{2} \equiv \frac{J_{\eta}^{2} \bar{B}_{0}^{2}}{R_{0}^{2}}, \quad \hat{\rho}_{m 0}=\frac{\rho_{m 0}}{\bar{\rho}_{m 0}}, \quad \hat{B}_{0}=\frac{B_{0}}{\overline{B}_{0}}, \quad \hat{\kappa}_{g}=\kappa_{g} R_{0}.
\end{equation}
Here, \(v _ { A 0 }\) is the Alfvén velocity, $\rho_{m0}$ is the equilibrium mass density, $B_{0}$ is the equilibrium magnetic field, \(\kappa_{g}\) is the geodesic curvature, $R_{0}$ is the radial position of the magnetic axis and all the quantities marked with a bar are calculated as reference values. \texttt{FALCON} code solves this linear system of second order coupled differential equations within different levels of simplification and calculates the relative SAW-ISW continuous spectra as described in \cite{falessi2019shear}. In particular, when the first and the second terms on the LHS of the second equation of (\ref{eq:366}) are negligible compared to the third one, i.e. when $\hat{\rho}_{m 0}/ \Gamma \beta \gg 1$, the set of coupled differential equations reduces to:
\begin{equation}
\label{eq:2}
\left(\partial_{\eta}^{2}+\hat{J}_{\eta}^{2} \hat{\rho}_{m 0} \Omega^{2}- 2 \Gamma \beta \hat{J}_{\eta}^{2}\hat{B}_{0}^{2}\hat{\kappa}_{g}^{2}\right) g_{1} = 0;
\end{equation}
describing the \emph{slow sound approximation} \citep{chu92}. Considering the limit $\hat{\rho}_{m 0}/ \Gamma \beta \rightarrow \infty$ allows neglecting the third term of Eq. (\ref{eq:2}); thus, recovering the incompressible ideal MHD limit. In the following work, the SAW-ISW continuum is solved within these three levels of progressive simplification that is, the whole Eqs. (\ref{eq:366}), the \emph{slow sound approximation} (\ref{eq:2}), and the incompressible ideal MHD limit.

Equations (\ref{eq:366}) and (\ref{eq:2}), are linear ODEs with periodic coefficients and, as extensively discussed in \cite{falessi2019shear}, Floquet theory can be applied. After re-writing Eqs. (\ref{eq:366}) and (\ref{eq:2}) as first order systems, it can be shown that they must have solutions in the following form: $\mathbf{x} =e ^ {i \nu \eta } \mathbf { P } ( \eta )$, where \(\mathbf{P}\) is a \(2 \pi\)-periodic vector function and $\nu$ is the characteristic Floquet exponent labelling the particular solution. In ``Floquet stable regions'', i.e. where $\nu$ is real, solutions are defined by radial singular structures corresponding to the SAW-ISW continuous spectrum. Meanwhile, unstable regions with complex $\nu$ are characterized by regular solutions in the frequency gaps. It is possible to find these solutions for any given value of $r$; thus, calculating the dispersion curves $\nu = \nu(\Omega,r)$, which involve only local quantities and describe wave packets propagating along magnetic field on a given flux surface. We emphasize the fact that one of the key benefits of this approach is that the dispersion curves are independent of the toroidal mode number, thus allowing high $n$ values to be analyzed without any resolution problems. Following \citet{chen16,zonca14a,zonca14b,chen17,falessi2019shear}, it is possible to relate the characteristic exponent to the toroidal and poloidal mode numbers; i.e., to express the continuous spectrum in the form:
\begin{equation}
\label{eq:4}
\nu ^ { 2 } ( \Omega , r ) = ( n q ( r ) - m ) ^ { 2 }.
\end{equation}
Integrating Eqs. (\ref{eq:366}) and (\ref{eq:2})  for different $\Omega$ values it is possible to calculate the continuous frequency spectrum for every value of the toroidal mode number retaining the effect of all poloidal harmonic couplings. Extensions to 3D/stellarator geometry and gyrokinetic theory would be obtained in the same way from the more general governing equations \citep{zonca14a,zonca14b,chen2016physics}. 

\begin{figure}
\centering
\includegraphics[width=.9\linewidth]{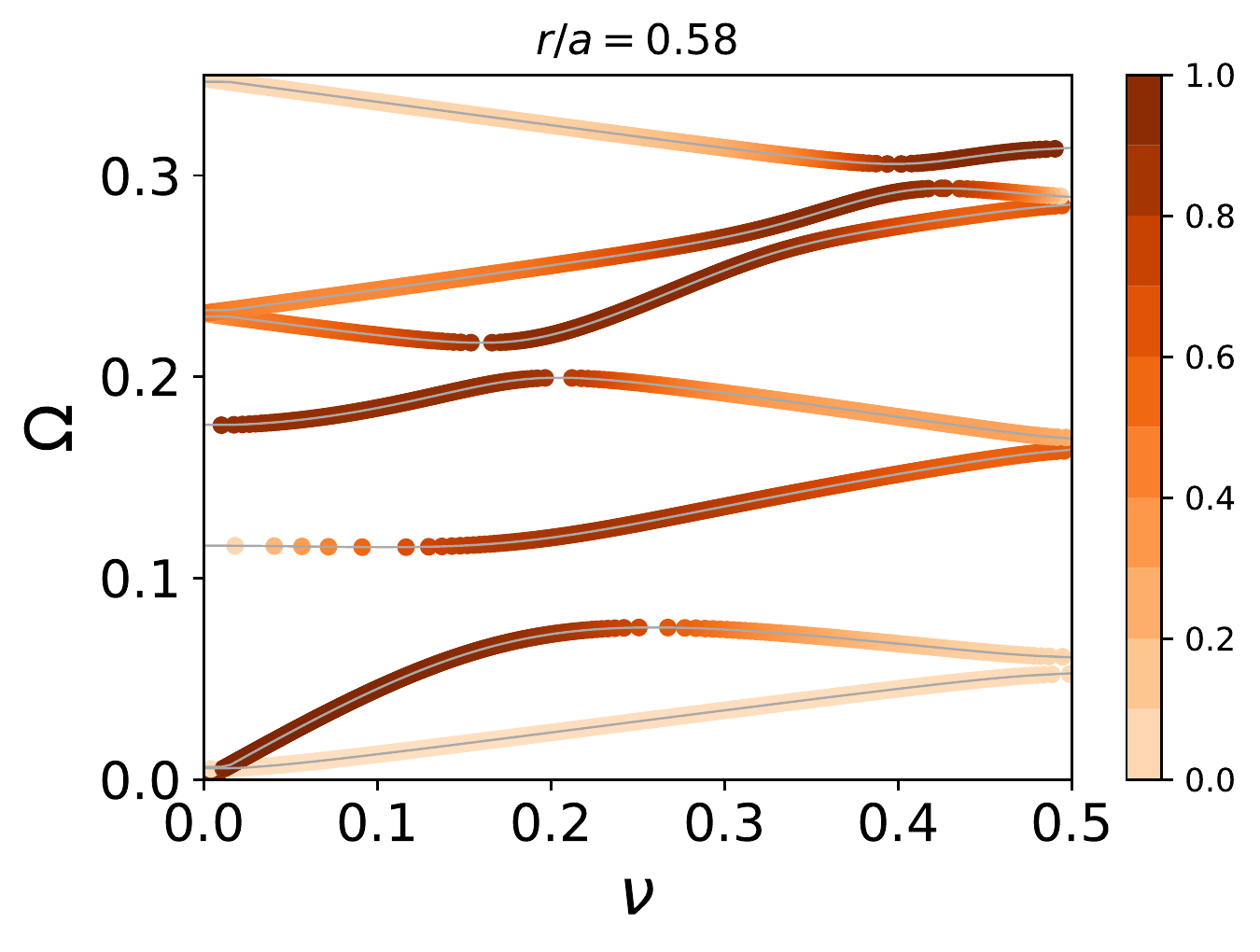}
\caption{\label{fig:org42bb9fee} Plot of the local dispersion curve $\nu(\Omega,r)$ at $r/a=0.58$ on DTT. The color bar show the Alfvénicity.}
\end{figure}

In \cite{chavdarovski09,zonca10,chavdarovski14}, the crucial role of polarization is emphasized, due to its fundamental role in determining the finite parallel electric field and, ultimately, the mode damping due to collisionless dissipation. The actual assessment of mode damping requires gyrokinetic theory to be employed, especially in high-$\beta$ plasmas due to diamagnetic effects which can significantly alter collisionless damping and cause the coupling of different branches.  In particular it is shown how fluctuations in the ``BAAE frequency'' range can have Alfvénic polarization, thus showing strongly reduced Landau damping. Given the profound connection of mode damping with the acoustic rather than the Alfvénic character of the mode, it is, therefore, of crucial importance to fully characterize the mode polarization even in the ideal MHD limit. This information, can be used as a proxy to anticipate the possible impact of kinetic effects, on the one hand. Furthermore, polarization knowledge allows to assess the actual coupling of a given (physical) fluctuation with the continuous spectrum, characterized by the corresponding polarization. Thus, in \cite{falessi2019shear}, without solving for the self-consistent mode structure and dispersion relation, a qualitative information about the effective role of resonant excitation of continuum structures is obtained from the so-called Alfvénicity parameter:
\begin{equation}
\label{eq:3}
{\cal A} = \frac{\int (g_1^{(i)}(\eta ; \nu,r))^2 d \eta}{\int [ (g_1^{(i)}(\eta; \nu, r))^2+(g_2^{(i)} (\eta; \nu, r))^2] d \eta}.
\end{equation}
In the considered MHD limit, ${\cal A} \simeq 1$ for SAW continuum while ${\cal A} \sim {\cal O}(\beta)$ for the acoustic continuum. A SAW polarized fluctuation has in general weak interaction with the acoustic polarized continuum. Meanwhile, an ISW polarization weakly interacts with the SAW continuous spectrum. Polarization and Alfvénicity, connected to each other as discussed by \cite{falessi2019shear}, are therefore of particular importance since the presence of continuum structures does not automatically imply damping \citep{chen74b,chen74a,hasegawa74}. As an illustrative example, in Fig. \ref{fig:org42bb9fee} we show the local dispersion curve $\nu(\Omega,r)$ obtained integrating Eqs. (\ref{eq:366}) on a given flux surface of the DTT reference scenario \citep{albanese17,albanese19} using the Alfvénicity as color bar. The usual MHD accumulation point at $\Omega = 0$ is the merging point of the low-frequency MHD fluctuations including inertia renormalization, characterized by predominant Alfvénic polarization \citep{chen17}. $\Omega = 0$ is also the accumulation point for the acoustic continuum with nearly orthogonal polarization with respect to the former one. Meanwhile, here, the BAAE gap is located at $\Omega \simeq 0.1$, while the BAE gap is at $\Omega \simeq 0.2$. It is instructive to note that polarization can be significantly modified along the dispersion curves, and that frequency by itself is not sufficient to fully characterize the properties (e.g., damping) of fluctuations, which also require detailed knowledge of spatial structures to be assessed.
\section{Application of \texttt{FALCON} to a DTT scenario}
\label{sec:orge40f96e}
\begin{figure}
\centering
\includegraphics[width=.45\linewidth]{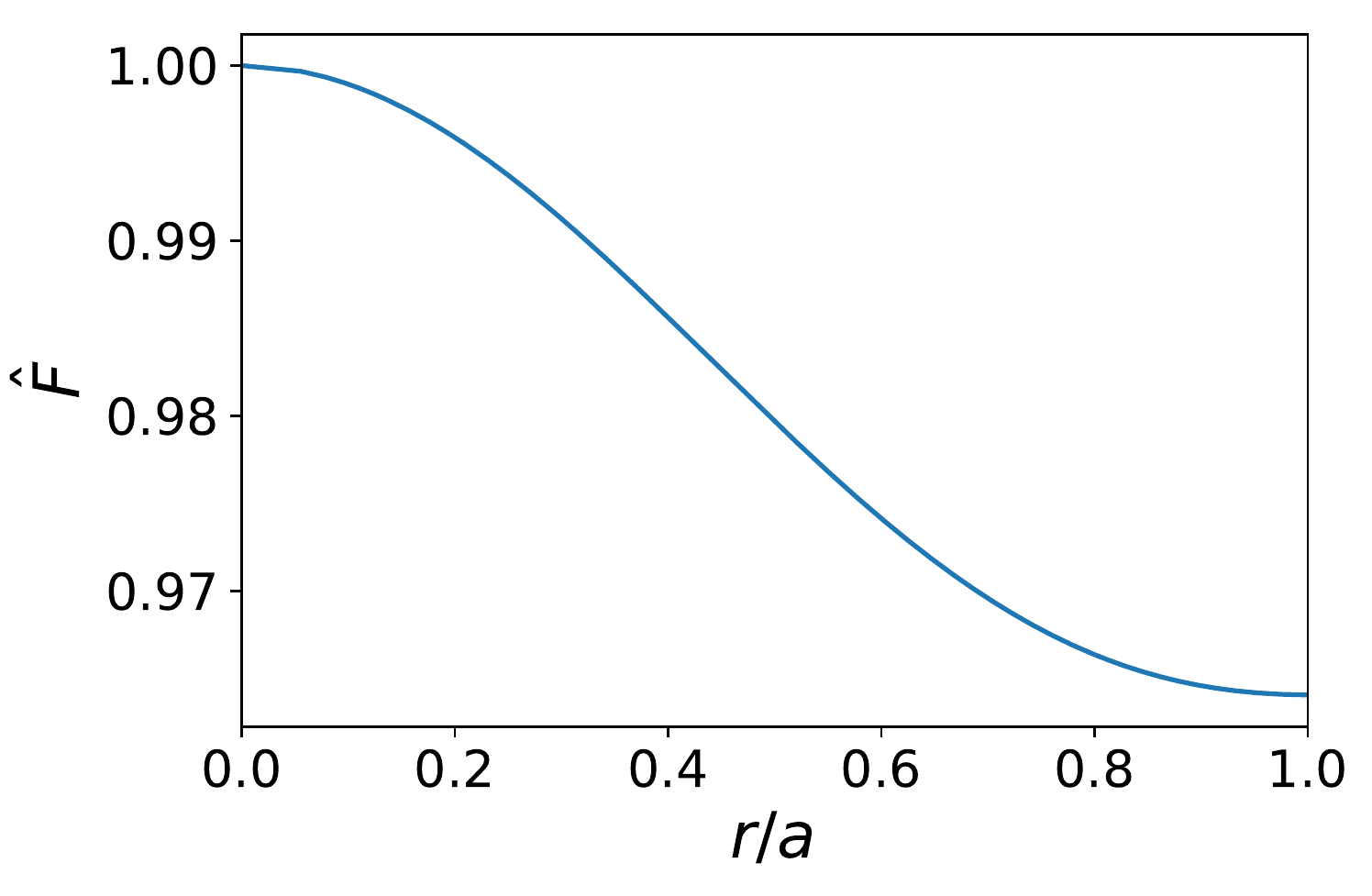}\includegraphics[width=.394\linewidth]{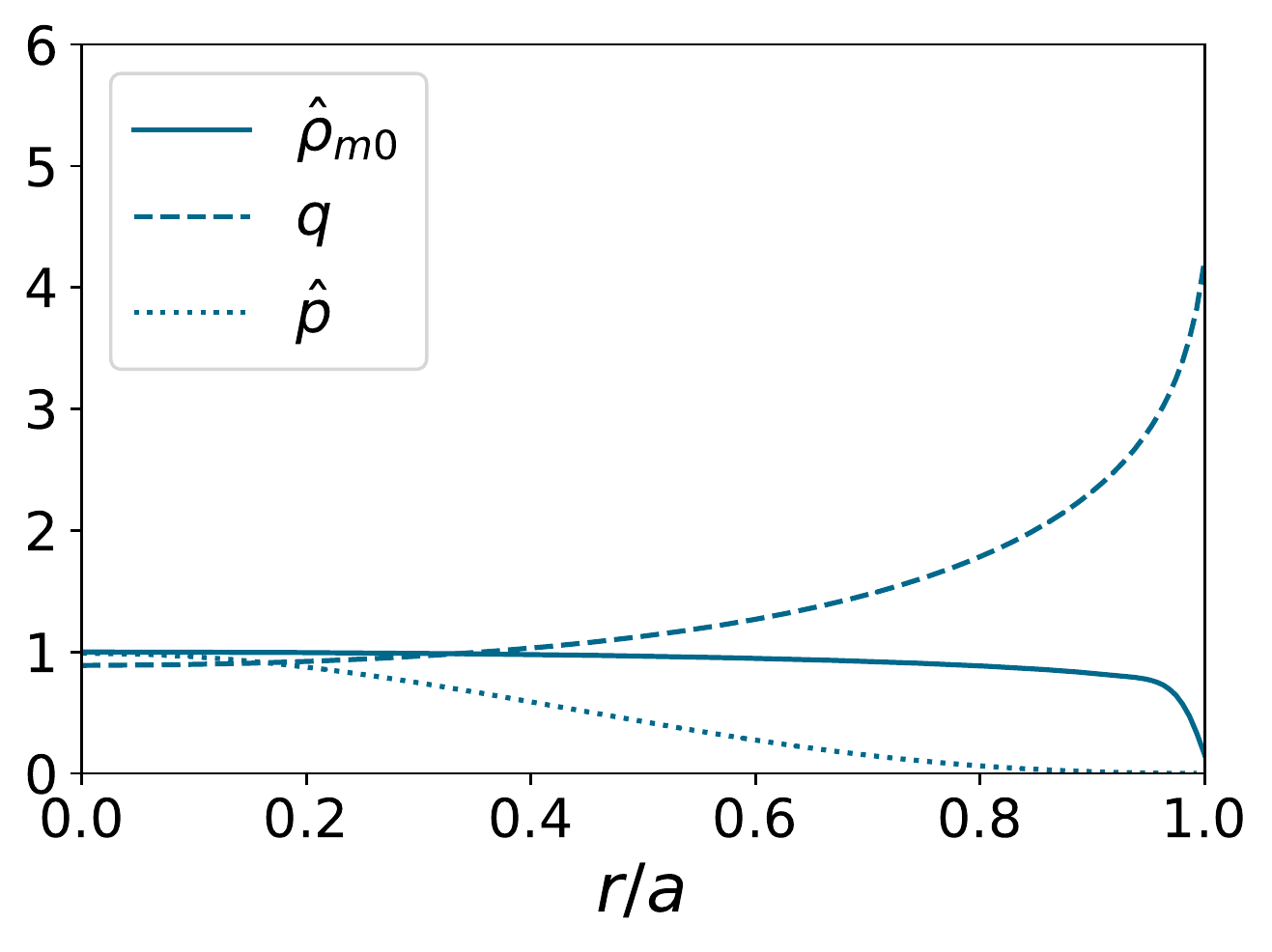}
\caption{Plots of the main profiles for the DTT reference scenario. Every quantity except $q$ is normalized to its value on the magnetic axis. In the left panel, the toroidal flux function is depicted while the right panel shows the kinetic pressure, the density and the safety factor $q$.}
\label{fig:dttprofiles}
\end{figure}

In this Section, in order to illustrate an application of \texttt{FALCON}, we calculate the frequency continuum of SAW and ISW waves in a DTT reference scenario \citep{albanese17,albanese19}. The same scenario has been analyzed in \citet{falessi2019shear}. The magnetic equilibrium has been originally calculated by means of the free boundary equilibrium evolution code CREATE-NL \citep{CREATE-NL} and further refined using the high-resolution equilibrium solver CHEASE \citep{CHEASE}. We consider a double null configuration, whose basic profiles are depicted in Fig. \ref{fig:dttprofiles} as a function of the normalized toroidal radius \(r/a\), i.e. the radial coordinate proportional to the square root of the toroidal magnetic flux. We note that, for the analyzed case, the kinetic pressure on axis is \(1.0768\times10^{6}~\mathrm{Pa}\), the flux function \(F\)  on axis is $12.95~\mathrm{Vs}$, while the density is $2.0739 \times 10\textsuperscript{19}~\mathrm{m}\textsuperscript{-3}$. The normalized density profile has been obtained by studies of plasma scenario formation using the fast transport simulation code METIS \citep{METIS}.

\begin{figure}
\centering
\includegraphics[width=.9\linewidth]{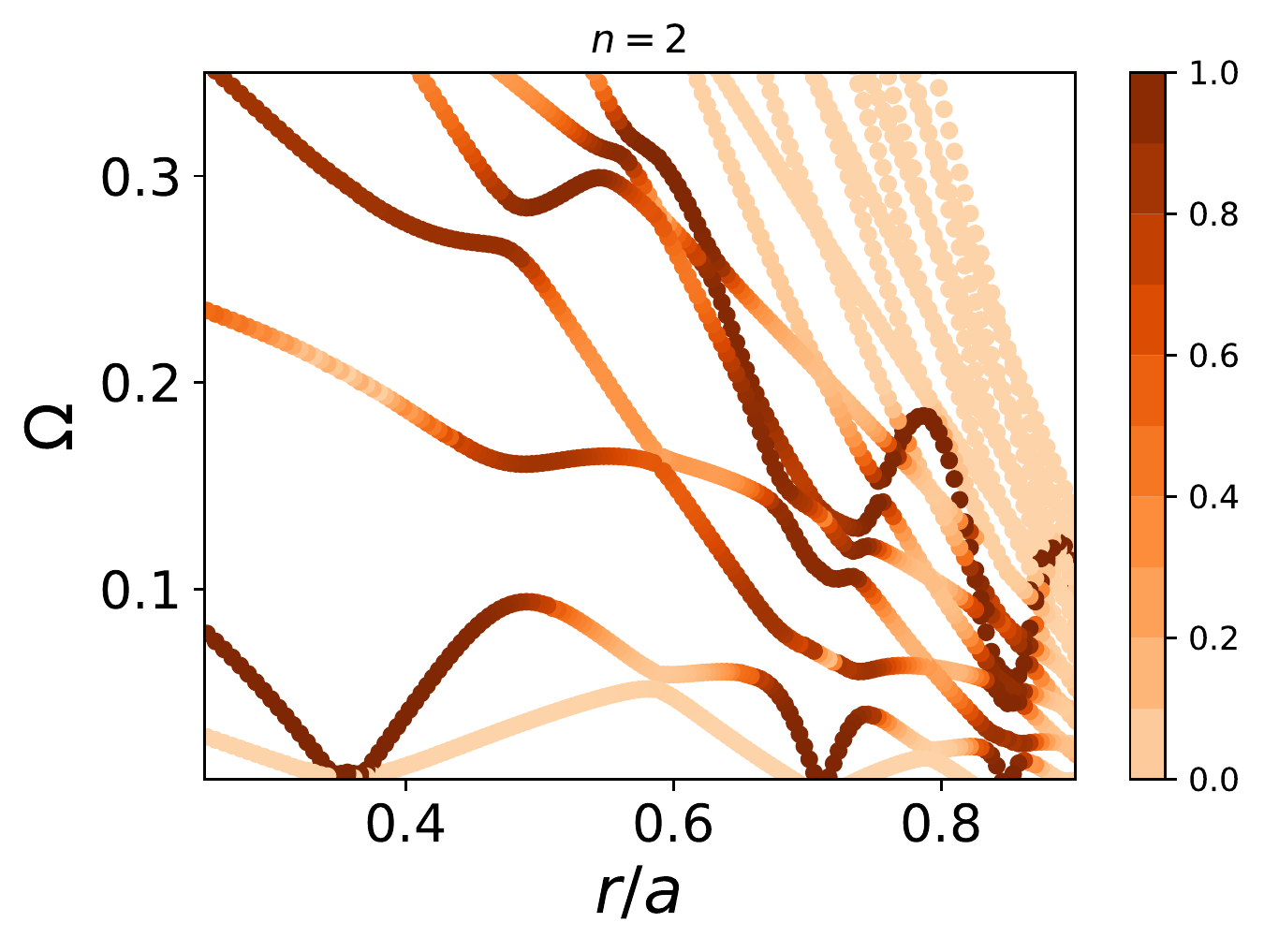}
\caption{\label{fig:org42bb9fe}SAW-ISW continuous spectrum as a function of \(r/a\) for \(n=2\) in DTT. The color bar show the Alfvénicity. For the sake of clarity, we did not plot all the acoustic branches nearby the plasma edge.}
\end{figure}

In Fig. \ref{fig:org42bb9fe}, we show the behavior of SAW-ISWs continuous spectra for a value of the toroidal mode number \(n = 2\). The code computes the Alfvénicity as introduced in Eq. (\ref{eq:3}), which is plotted as a color-bar. In particular, we note that this allows to isolate SAW polarized branches when approaching the plasma edge. In fact, in this region,  high order ISW side-bands are present due to the low value of plasma pressure and coupled because of the shaping of the equilibrium. However, most of them are not physically relevant due to the stronger kinetic damping of the ISW branch. Furthermore, due to their negligible Alfvénicity, they would be weakly coupled to any fluctuation with prevalent Alfvénic polarization, for which the ``web of acoustic continua'' near the plasma boundary is nearly transparent.

\begin{figure}
\centering
\includegraphics[width=.9\linewidth]{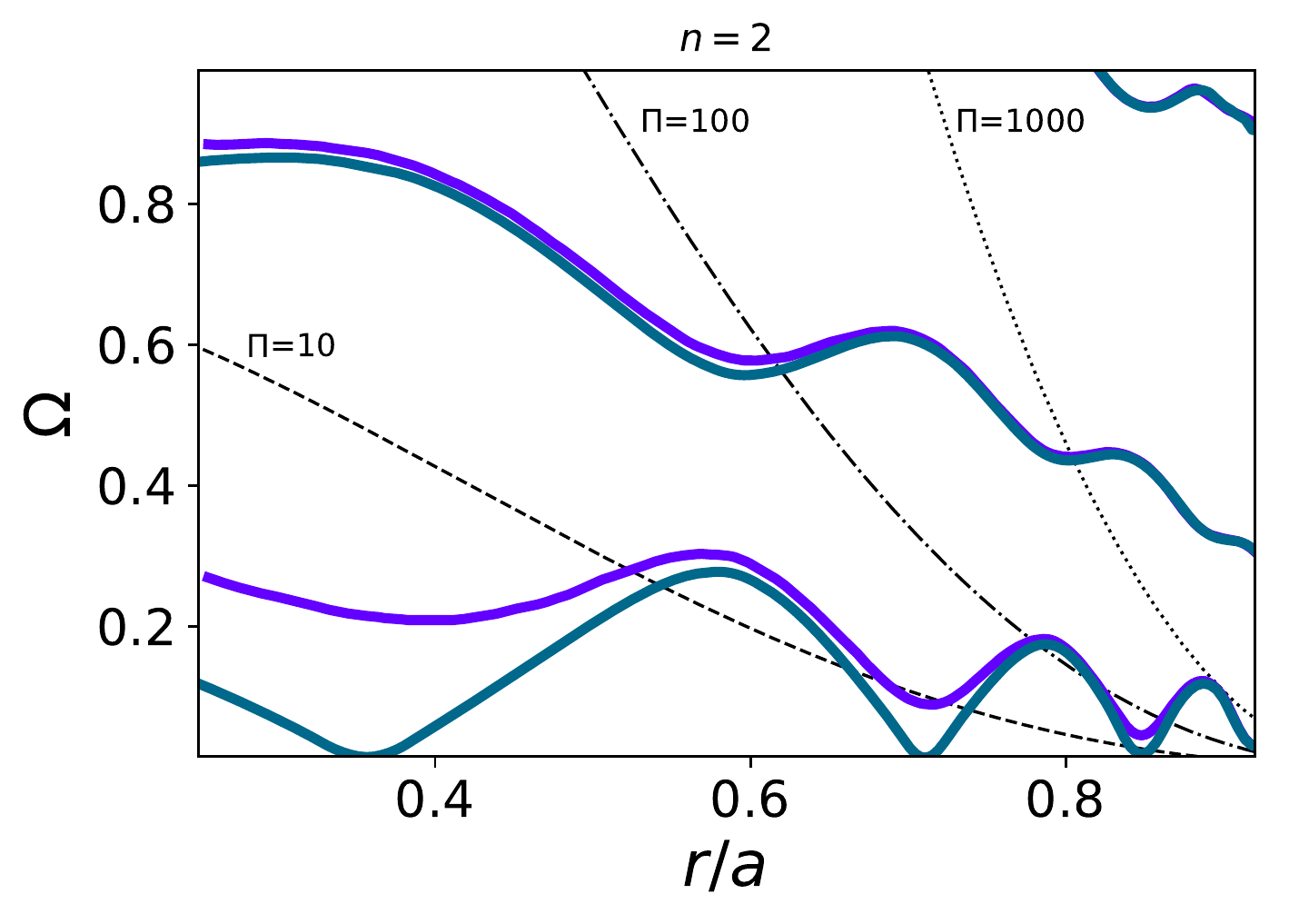}
\caption{\label{fig:org42624f8}SAW-ISW continuous spectrum calculated using the slow-sound  approximation as a function of \(r/a\) for \(n=2\) on DTT is represented in purple. Results obtained in the incompressible ideal MHD limit are plotted in blue. Dotted and/or dashed lines describe curves of constant \(\Pi = \langle\hat{\rho}_{m 0}/ \Gamma \beta \rangle_{\psi}\).}
\end{figure}

\begin{figure}
\centering
\includegraphics[width=.9\linewidth]{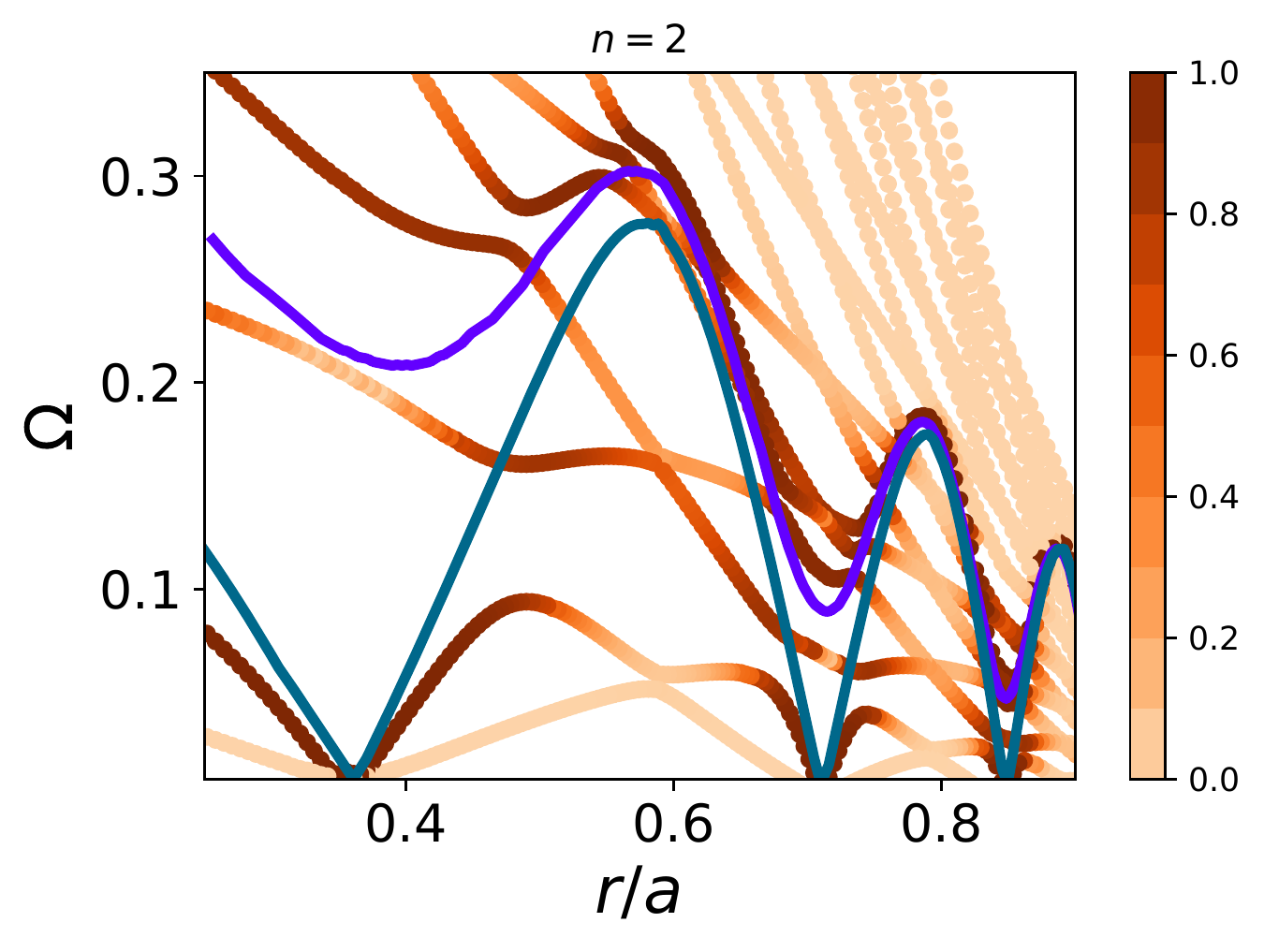}
\caption{\label{fig:org0afb4bb}SAW-ISW continuous spectrum obtained with different level of simplifications on DTT as a function of \(r/a\) for \(n=2\). The first branches obtained, respectively, with the \emph{slow sound approximation} and ideal MHD incompressible limit are marked in purple and blue. The color bar shows the Alfvénicity.}
\end{figure}

As already mentioned, \texttt{FALCON} solves the SAW-ISW continuum spectrum within different levels of simplification. In particular, in Fig. \ref{fig:org42624f8}, we show the results obtained by using the \emph{slow sound approximation}, i.e. Eq. (\ref{eq:2}), and compare it to the results obtained neglecting the coupling with ISWs in the incompressible ideal MHD limit. In the same Figure, we plot the isolines of the flux surface averaged quantity \(\Pi = \langle\hat{\rho}_{m 0}/ \Gamma \beta \rangle_{\psi}\). As expected, the agreement between the two approximations is quite good when \(\Pi\) is larger, i.e. approaching the plasma edge and at higher frequencies. By direct inspection of Fig. \ref{fig:org42624f8}, we see that the first branch of the ideal SAW continuous spectrum is for a substantial part under the first $\Pi$ isoline, i.e. $\Pi = 10$. In this region SAW-ISW corrections with respect to the incompressible ideal MHD case are consistent as we can infer from Fig. \ref{fig:org42bb9fe}. In particular the BAAE gap \citep{gorelenkov09,gorelenkov07a,gorelenkov07b} extends up to $\Omega \sim 0.25$ and, therefore, studying the complete system, i.e. Eq. (\ref{eq:366}), is mandatory in this range of frequencies. The \emph{slow sound approximation} seems to be adequate in the region between the first two  isolines, i.e. $10<\Pi<100$ since the correction are very small with respect to the whole system. As anticipated, the incompressible MHD continuous spectrum agrees well with this approximation in the remaining portion of the plot.  We note that, nearby the plasma edge, $\Pi$ increase very fast with respect to $\Omega$ variations. For this reason, results obtained with the three different level of approximation are becoming more and more similar as the frequency decreases. This can be clearly seen in Fig. \ref{fig:org0afb4bb}. In principle, the contour plot of $\Pi$ could be used routinely by the code to automatically choose the appropriate reduced model in the continuum equations, if needed, to possibly reduce the computational load and to illuminate the relevant physical processes. 
In Fig. \ref{fig:org0afb4bb}, we compare these results with the SAW-ISWs continuous spectrum already shown in Fig. \ref{fig:org42bb9fe}.  Again, the agreement between the \emph{slow sound approximation} and the respective Alfvénic branch of the complete system is quite good for high frequencies and approaching the plasma edge. Furthermore, in regions where \(\Pi\) is significantly high the ISWs coupling can be even neglected and the incompressible ideal MHD limit can be used. As already stated, complex low frequencies continuum structures, e.g. the frequency gaps due to SAW-ISW and/or ISW-ISW couplings dubbed BAAE gap can be described only by means of the complete SAW-ISW system.
\section{Application of \texttt{FALCON} to an AUG scenario}
\label{sec:org9b666dd}
\begin{figure}
\centering
\includegraphics[width=.45\linewidth]{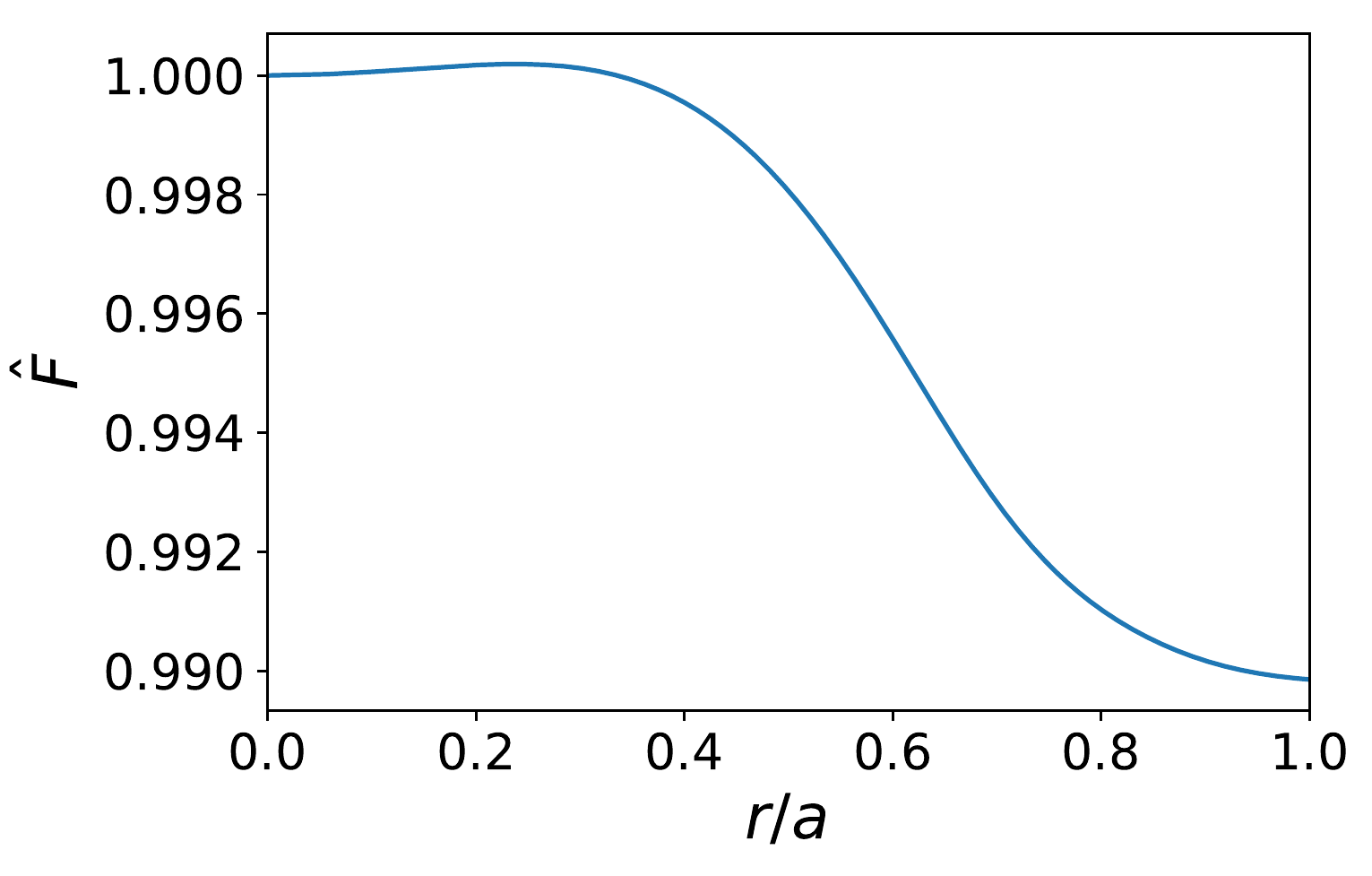}\includegraphics[width=.394\linewidth]{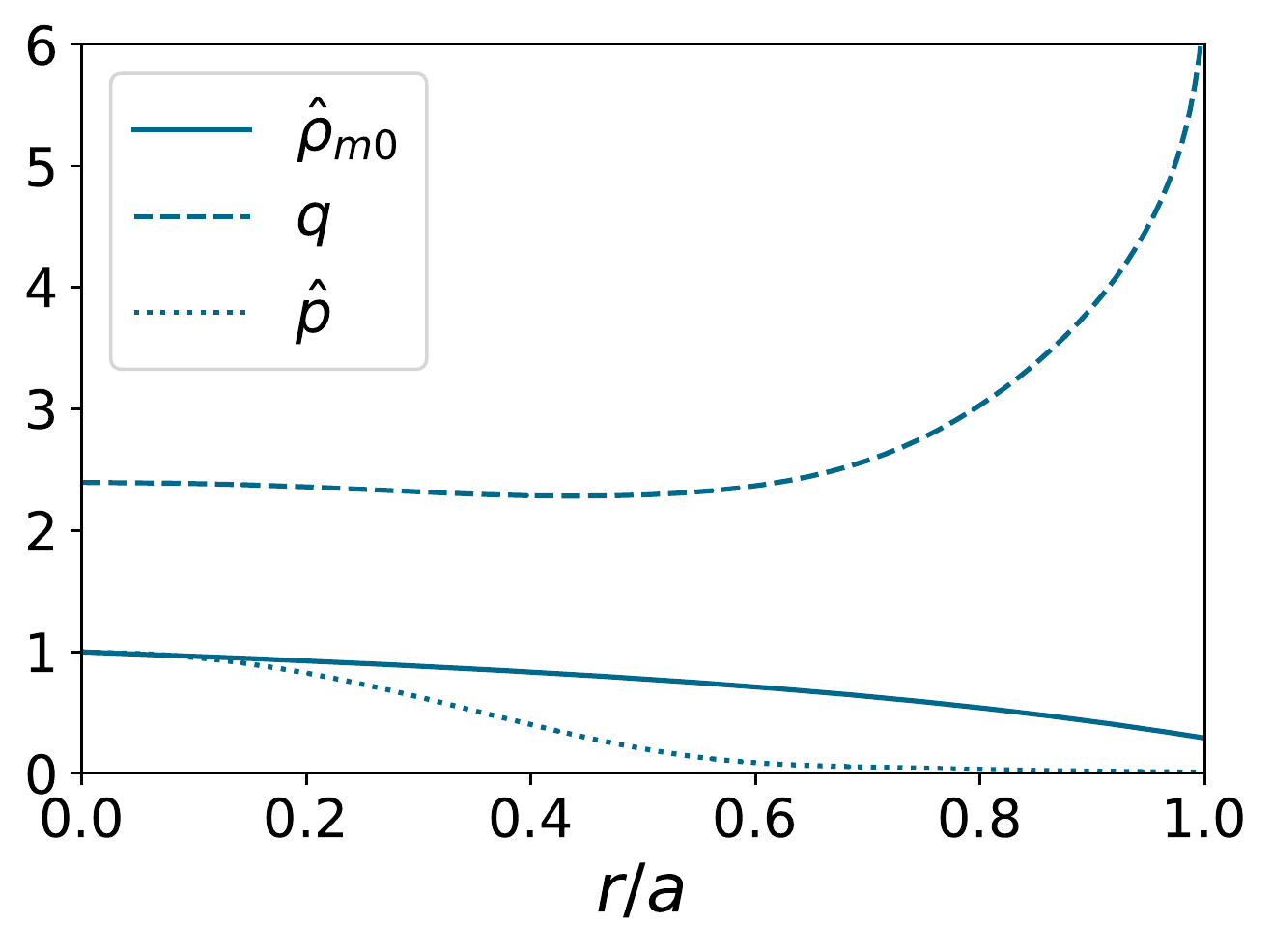}
\caption{Plots of the main profiles in the AUG \(\#31213\) discharge. Every quantity except $q$ is normalized to its value on the magnetic axis. In the left panel, the toroidal flux function is depicted, while the right panel shows the kinetic pressure, the density and the safety factor $q$.}
\label{fig:augprofiles}
\end{figure}

In this Section, we perform the same analysis of Sec. \ref{sec:orge40f96e} applied to the AUG \(\#31213\) discharge. The magnetic equilibrium has been originally calculated by means of HELENA \citep{HUYSMANS_1991} and further refined using the high-resolution equilibrium solver CHEASE \citep{CHEASE}. We consider a single null scenario whose basic profiles are shown in Fig. \ref{fig:augprofiles} as a function of the normalized toroidal radius \(r/a\). We note that, for the analyzed case, the kinetic pressure on axis is \(3.0781\times10^{4} ~\mathrm{Pa}\), the flux function \(F\)  on axis is $3.68~\mathrm{Vs}$, while the density is $1.3607 \times 10 \textsuperscript{19}\mathrm{m\textsuperscript{-3}}$. It should be noted that this specific AUG plasma has a particularly low plasma $\beta$ in order to maximise the ratio $\beta_{EP}/\beta \sim 1$ for the detailed study of EP-induced instabilities \citep{Horv_th_2016}. It is chosen here to demonstrate the influence of low and high $\beta$ on the polarization.

\begin{figure}
\centering
\includegraphics[width=.9\linewidth]{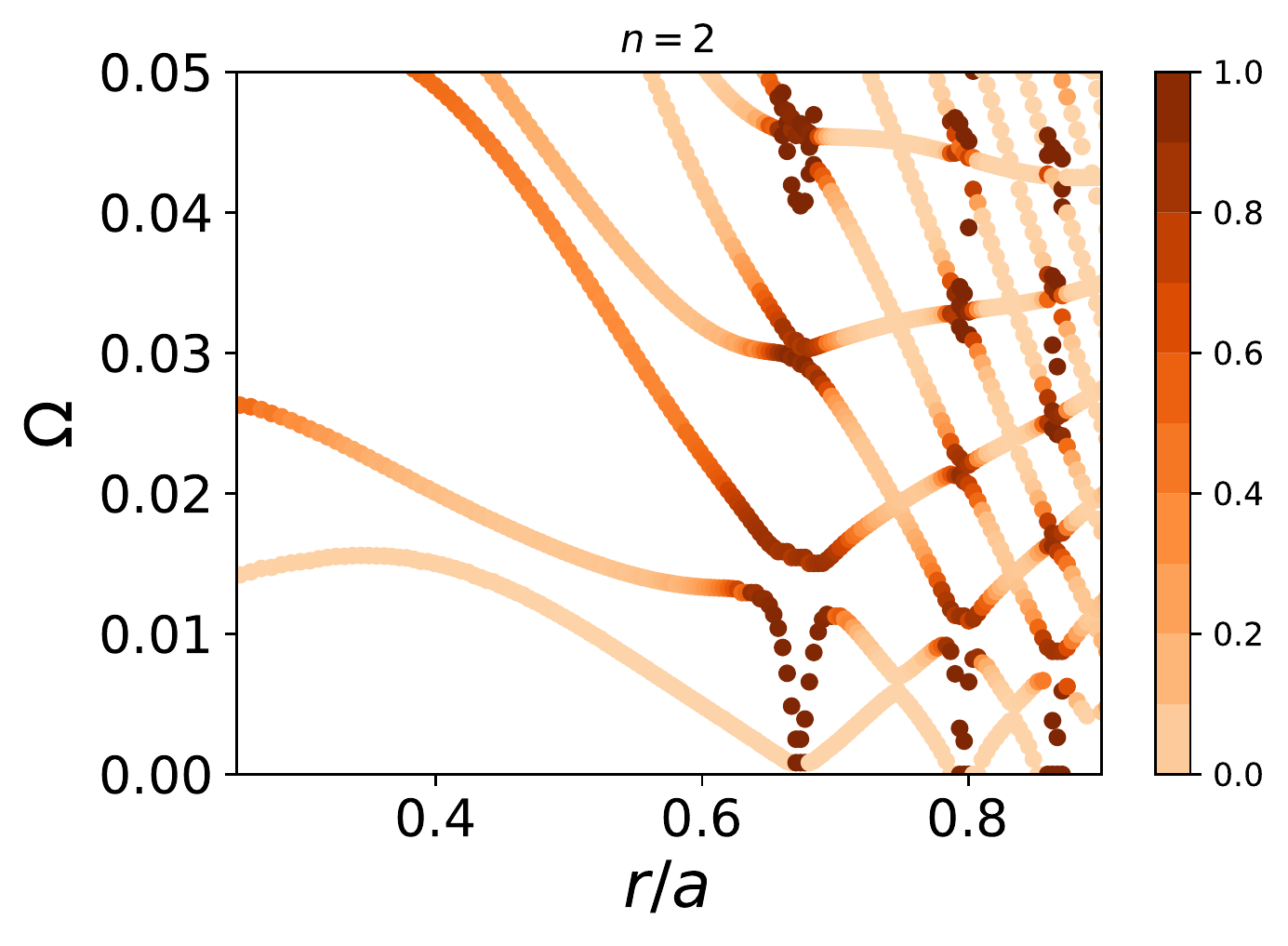}
\caption{\label{fig:orgcd29a35}SAW-ISW continuous spectrum as a function of \(r/a\) for \(n=2\) in the AUG \(\#31213\) discharge. The color bar shows the Alfvénicity.}
\end{figure}

\begin{figure}
\centering
\includegraphics[width=.9\linewidth]{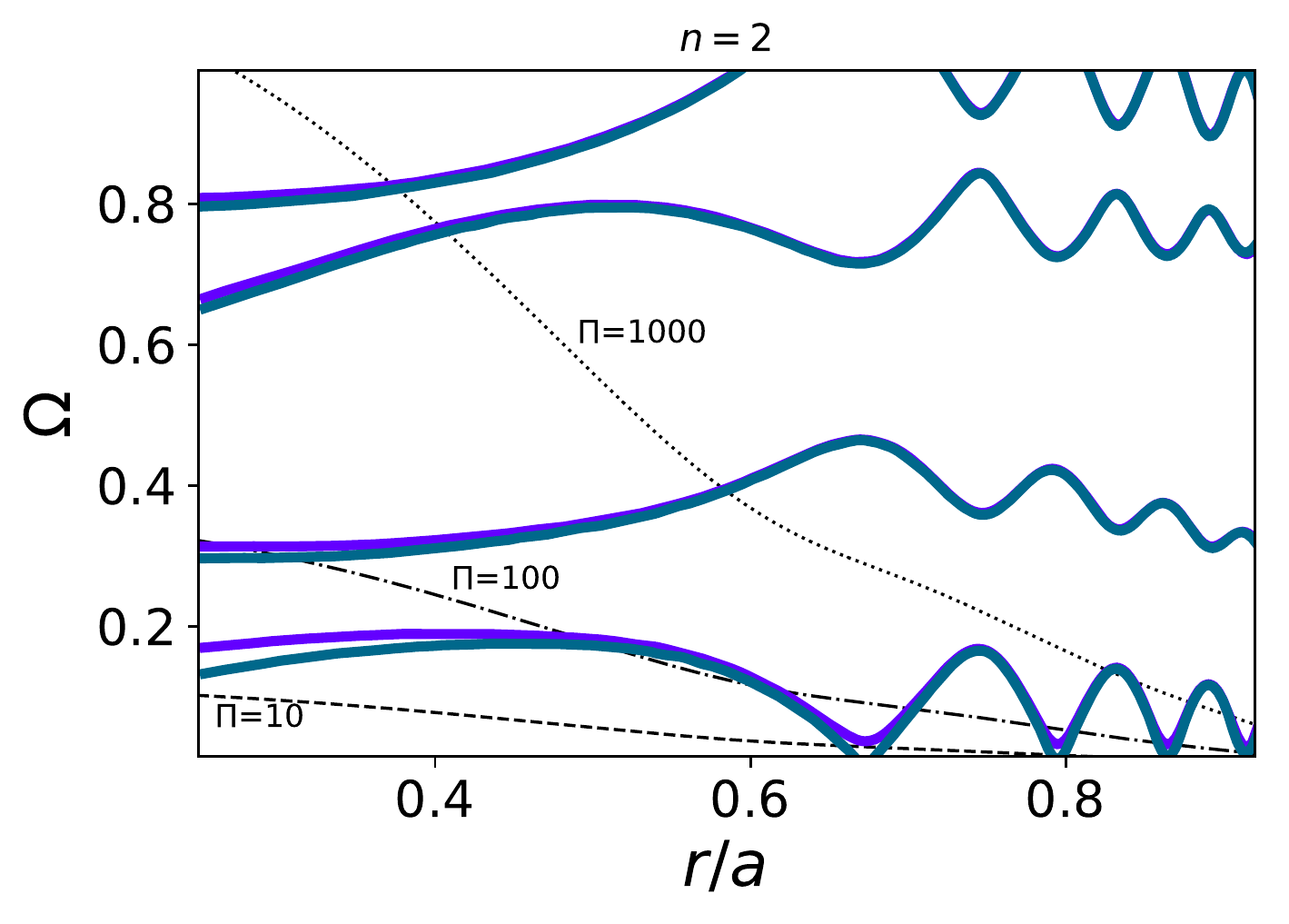}
\caption{\label{fig:org16bc935}SAW-ISW continuous spectrum in the AUG \(\#31213\) discharge calculated using the \emph{slow sound approximation} as a function of \(r/a\) for \(n=2\) is represented in purple. Results obtained in the incompressible ideal MHD limit are plotted in blue. Dotted and/or dashed lines describe curves of constant \(\Pi = \langle\hat{\rho}_{m 0}/ \Gamma \beta \rangle_{\psi}\) .}
\end{figure}

In Fig. \ref{fig:orgcd29a35} we show the results for the obtained SAW-ISWs spectra. The multiplicity of ISW branches with respect to the DTT reference case is due to the lower $\beta$. In particular, the BAAE gap extend up to $\Omega \sim 0.06$ As in the DTT case, Alfvénicity allows to distinguish SAW polarized branches. Results obtained using the \emph{slow sound approximation} and incompressible ideal MHD limit are illustrated in Fig. \ref{fig:org16bc935}. As expected, the region of validity of this approximation in the \((\Omega, r/a)\) plane is increased with respect to the DTT reference case, confirming that Alfvénic fluctuations in DTT plasmas will be more importantly affected by kinetic and compressibility effects than AUG \citep{lauber08,lauber09,lauber12,lauber2013,Bierwage2017}. The \emph{slow sound approximation} correctly describes most of the propagation of SAW-ISW waves with frequencies above the BAE gap while Alfvénic branches are correctly represented in the majority of the plot also by the incompressible ideal MHD limit. The structures at the BAAE frequency range  appear to be less Alfvénic polarized with respect to the DTT reference scenario. Again, this is consistent with the findings reported in the literature \citep{lauber08,lauber09,lauber12,lauber2013,Bierwage2017}, which show that the acoustic branch in AUG is typically strongly damped and that modes in the BAAE frequency range are essentially kinetic ballooning modes with predominant Alfvénic polarization, where diamagnetic effects play important roles. More precisely, damping of the BAAE branch at AUG is found to be significantly reduced only very close to rational surfaces, again consistently with theoretical predictions \citep{lauber2013}.

\section{Conclusions}
\label{sec:org5e2748a}
In this work, following \citet{falessi2019shear}, we have calculated the SAW-ISW continuous spectrum in realistic Tokamak geometry by means of the newly developed \texttt{FALCON} code, using the theoretical framework introduced in \citet{chen16,zonca14a,zonca14b,chen17} based on the mode structure decomposition/ballooning formalism \citep{lu12}. Apart from its simplicity, the main advantages of this formulation have been reviewed with particular attention to the ability to isolate the radially local (singular) behaviors of the continuous spectra and handle fluctuations with large toroidal mode numbers efficiently. These motivations guided us to create a new computational tool, i.e. the \texttt{FALCON} code. The proposed methodology clearly shows the significance of fluctuation polarization in the evaluation of damping by resonant absorption of singular radial continuous spectrum structures. Therefore, following \cite{falessi2019shear}, an Alfvénicity parameter is introduced, which gives a qualitative estimate of the coupling strength of Alfvénic fluctuations to the acoustic continuum without solving for the actual fluctuation structures. Adopting DTT and AUG reference scenarios as example of, respectively, high- and low-$\beta$ tokamak equilibria, the Alfvénicity parameter has been computed in the $(\Omega,r/a)$ plane, thus allowing us to properly characterize the physical nature of the observed spectra. Meanwhile, the \texttt{FALCON} code can now operate with different levels of simplification, if needed for further speed-up, and, in particular, use the \emph{slow sound approximation}, which is shown to be particularly relevant for the study of relatively low $\beta$ equilibria. Moreover, this approach not only allows one to address the structures of SAW and ISW continuous spectra in general toroidal geometry, but also provides the most straightforward way of computing the generalized inertia in the GFLDR for generic Alfvén and Alfvén acoustic modes \citep{chen16,zonca14a,zonca14b}. The current formalism, based on Floquet theory, is extremely general and could be extended to $3D$ plasma equilibria such as stellarators and/or to the kinetic description of coupled SAW and ISW continua. Finally, it could be applied to calculate the boundary conditions for mode structures and their relative dispersion relations consistent with the GFLDR.

\section{Acknowledgment}
\label{sec:orgf89e966}
This work has been carried out within the framework of the EUROfusion Consortium and has received funding from the Euratom research and training programme 2014-2018 and 2019-2020 under grant agreement No 633053. The views and opinions expressed herein do not necessarily reflect those of the European Commission.

Authors thank the DTT and the ASDEX Upgrade Teams for providing the information on the respective reference scenarios.
\bibliographystyle{jpp}
\bibliography{jppbib}
\end{document}